\documentclass[prl,superscriptaddress,showpacs,twocolumn]{revtex4}
\usepackage{graphicx}

\usepackage{amsmath,amssymb,latexsym,amsfonts}
\begin{document}
\title{Tunable Fano Resonances in Transport through Microwave Billiards}
\author{S.~Rotter}
\email{rotter@concord.itp.tuwien.ac.at}
\author{F.~Libisch}
\author{J.~Burgd\"orfer}
\affiliation{Institute for Theoretical Physics, Vienna University of
Technology, A-1040 Vienna, Austria}
\author{U.~Kuhl}
\author{H.-J. St\"ockmann}
\affiliation{Fachbereich Physik,
Philipps-Universit\"at Marburg, D-35032 Marburg, Germany}
\date{\today}

\begin{abstract}
We present a tunable microwave scattering device that allows the
controlled variation of Fano line shape parameters in transmission
through quantum billiards. Transport in this device is nearly fully
coherent. By comparison with quantum calculations, employing the
modular recursive Green's-function method, the scattering
wave function and the degree of
residual decoherence can be determined. The parametric variation
of Fano line shapes in terms of interacting resonances is
analyzed.
\end{abstract}

\maketitle

Asymmetric Fano line shapes are an ubiquitous feature of resonance
scattering when (at least) two different pathways connecting the
entrance with the exit channel exist. Fano resonances have been
observed in a wide array of different subfields of physics starting
with photoabsorption in atoms \cite{beut,fano35,fano61}, electron
and neutron scattering \cite{adair,simpa63}, Raman
scattering \cite{cer73}, photoabsorption in quantum well
structures \cite{feist97}, scanning tunnel microscopy \cite{mad98},
and ballistic transport through quantum dots
(``artificial atoms'') \cite{goeres00,noeckel94,rotter2,koba}. Interest in
observing and analyzing Fano profiles is driven by their high
sensitivity to the details of the scattering process.
For example, since Fano parameters reveal the presence and the nature
of different (non) resonant pathways, they can be used to
determine the degree of coherence in the scattering device.  This is
due to the fact that decoherence converts Fano resonances
into the more familiar limiting case of a
Breit-Wigner resonance. Furthermore, they provide detailed
information on the interaction between nearby resonances leading
to ``avoided crossings'' in the complex plane \cite{heiss91,burg95},
and to stabilization of discrete states in the continuum
(``resonance trapping'' \cite{ro,rott}).\\Exploiting 
the equivalence of the scalar Helmholtz equation for
electromagnetic radiation in cavities with conducting walls and
the Schr\"odinger equation subject to hard-wall boundary
conditions  \cite{stock}, we have designed a scattering device
(Fig.~\ref{fig:1}) that allows the controlled tuning of Fano resonances for
transport through quantum billiards. The evolution of the Fano
parameters as a function of the tuning parameter, in the present
case the degree of opening of the leads, can be traced in
unprecedented detail, since decoherence due to dissipation is kept at
a low level. By comparison with
calculations employing the modular recursive Green's function
(MRGM) \cite{rotter2,rotter}, the parametric variation of Fano resonances on
the degree of decoherence can be quantitatively accounted for.
Furthermore the relevant pathways can be unambiguously identified in terms of
wave functions representing the contributing scattering channels.
Due to the equivalence between microwave transport and single-electron
motion in two dimensions, our device also simulates ballistic
electron scattering through a  quantum dot. In contrast to recent
investigations of mesoscopic dots and single-electron 
transistors \cite{goeres00,koba,brems}, where comparison between theory
and experiment has remained on a mostly qualitative level, our
model system allows for a detailed quantitative analysis of all
features of tunable resonances.\\Our microwave scattering device
consists of two commercially available waveguides
(height $h$=7.8\,mm, width $d$=15.8\,mm, length $l$=200\,mm)
which were attached both to the entrance and
the exit side of a rectangular resonator
(height $H$=7.8\,mm, width $D$ = 39\,mm, length $L$ = 176\,mm).
At the junctions to the cavity metallic diaphragms of different
openings were inserted (Fig.~\ref{fig:1}). The microwaves with frequencies
between 12.3 and 18.0 GHz, where two even transverse modes are excited
in the cavity and one transverse mode in each of the leads,
are coupled into the waveguide via an adaptor to ensure strong 
coupling.\\The experimental results are compared with the predictions
of the MRGM. We solve the $S$ matrix for the single particle Schr\"odinger
equation for this ''quantum dot'' by assuming a constant potential
set equal to zero inside and infinitely high outside of a hard-wall boundary.
At asymptotic distances, scattering boundary conditions are
imposed in the leads.
The coupling of the leads to the cavity of length $L$ can be varied by two
diaphragms which are placed symmetrically at the two lead openings.
The lead width $d$ and the
width of the rectangular cavity $D$ determine how many flux-carrying
modes are open at a certain energy $\varepsilon$ in each of the
three scattering regions (lead-cavity-lead). We consider in the
following the range of wavenumbers 
%$k_2^{c} \leq k \leq k_3^{c}$
where one flux-carrying mode is open in each of the leads, while
the first and second even transverse modes are open inside the cavity, thus
providing two alternative pathways of quantum transport.
In order to further characterize the interfering paths, we decompose the
transport across the cavity into a multiple scattering series
involving three pieces \cite{wirt03}, each of which
is characterized by a  mode-to-mode
transmission (reflection) amplitude or a propagator:
(1) the transmission of the incoming flux from the
 left into the cavity, $t^{(L)}$, or reflection back into the
lead, $r^{(L)}$, (2) the propagation inside the cavity from the left
 to the right, $G^{(LR)}$, or from the right to the left,
 $G^{(RL)}$, and (3) the transmission from
 the interior of the cavity to the right, $t^{(R)}$,
 or internal reflection at each of the two vertical cavity
walls with amplitude $r^{(R)}$. For the Green's functions (i.e.~propagators)
$G^{(LR)}( x_R, x_L )$ and $G^{(RL)}(x_L,x_R)$
we choose a mixed representation which is local in $x$, and
 employs a spectral sum over transverse modes,
 $G^{(LR)}( x_R, x_L )=G^{(RL)}( x_L, x_R )=
 \sum_n\vert n \rangle
 \exp(ik_n |x_R-x_L|)\langle n \vert$,
where $x_{R,L}$ are the $x$-coordinates of the right (left) lead junction
with $|x_R - x_L|= L$. The longitudinal momentum for each
 channel $n$ in the cavity is given by $k_n=\sqrt{k^2-(k_n^c)^2}$,
with the momentum $k=\sqrt{2\varepsilon}$
and the threshold $k$-values $k_n^c=n\pi/D$. Decoherence due to
dissipation of the microwave power in the cavity walls can be
easily incorporated by analytically continuing $k_n$ into the complex plane,
$k_n=\sqrt{k^2-(k_n^c)^2}+i\kappa$.
The quantitative analysis of Fano resonances
for these systems can be used to accurately determine
the degree of dissipation present. The multiple scattering expansion
of the transmission amplitude $T$ is then given by
\begin{align}
\label{eqn:3}
 T (k) &=t^{(L)} G^{(LR)} \left\{
 \sum\limits_{n=0}^\infty [ r^{(R)}G^{(RL)} r^{(L)} G^{(LR)} ]^n
 \right\}\, t^{(R)}=\nonumber\\&=t^{(L)} G^{(LR)} \left[1- 
r^{(R)}\, G^{(RL)} r^{(L)} G^{(LR)}\right]^{-1} t^{(R)}\,.
 \end{align}
 The identification  of the resonant and non-resonant pathways with
help of Eq.\ (\ref{eqn:3}) is straightforward: due to the absence of
inter-channel mixing in the rectangular (i.e.~non-chaotic) cavity,
the non-resonant contribution corresponds to the $n = 0$ term of
the sum describing direct transmission while the resonant contribution
is made up by all multiple-bounce contributions $(n \geq 1 )$. 
The various amplitudes entering Eq.\ (\ref{eqn:3})
can be parameterized in terms of four phases and two moduli \cite{kim01}:
the modulus, $s$, of the reflection amplitude of the wave incoming
in mode 1 and reflected into mode 1 at the left diaphragm,
 \begin{equation}
 \label{eqn:4}
 r_{11}^{(L)} = s e^{i \phi_r}\,,
 \end{equation}
and the modulus, $p$, of the partial injection amplitude of the
incoming wave into the lowest mode of the cavity, corrected for
the partially reflected  flux,
\begin{equation}
\label{eqn:5}
t_{11}^{(L)} = t_{11}^{(R)} = p \sqrt{1-s^2}\,e^{i \phi_t^{(1)}} \, .
\end{equation}
Because of the symmetry of the scattering device, the injection
(ejection) amplitude at the left (right) side are equal. Accordingly,
 the injection amplitude into the second even mode of the cavity is given by
\begin{equation}
\label{eqn:6}
t_{12}^{(L)} = t_{21}^{(R)} = \sqrt{(1-p^2) (1-s^2)}\,e^{i \phi_t^{(2)}} \, .
\end{equation}
Analogous expressions can be deduced  \cite{kim01}
for the other partial amplitudes entering Eq.\ (\ref{eqn:3}). We
omit a detailed analysis of the phases in
Eqs.\ (\ref{eqn:4},\ref{eqn:5},\ref{eqn:6}) since they do not explicitly enter
our analysis in the following. The key observation in the present
context is that the square module $s^2$ is monotonically decreasing in
between the limiting values $s^2=1$ for zero diaphragm opening
($w=0$) and $s^2\approx 0$ for fully open diaphragms ($w=d$).
Inserting Eqs. (\ref{eqn:4},\ref{eqn:5},\ref{eqn:6}) into
Eq.\ (\ref{eqn:3}), a closed yet complicated expression for the
transmission probability $|T (\varepsilon, s)| $ 
as a function of the energy $\varepsilon$
and the opening parameter $s$ can be derived.
Close to a given resonance $\varepsilon_i^R$ this expression can
be approximated by the Fano form \cite{fano61,kim01},
\begin{equation}
\label{eqn:7}
|T (\varepsilon, s)|^2  \approx \frac{|\varepsilon - \varepsilon_i^R(s) +
q_i (s) \Gamma_i (s) /2|^2}{\left[ \varepsilon - \varepsilon_i^R (s) \right]^2
+ \left[ \Gamma_i (s)/2 \right]^2}\,,
\end{equation}
where $\varepsilon_i^R (s)$ is the position of the $i$-th resonance,
$\Gamma_i (s)$ its width, and $q_i (s)$ the complex Fano asymmetry parameter,
all of which depend on $s$. Window resonances appear in the limit
$q \rightarrow 0$ while the Breit-Wigner limit is reached for $|q| \gg 1$.
It should be noted that, in general, $q$ cannot be simply
identified with the ratio of resonant to non-resonant coupling
strength \cite{eich03,fried}. Fig.~\ref{fig:2} presents both the
experimental and theoretical dependence of the transmission
probability $|T|^2$ on $k $ (or $\varepsilon$). In the measurement,
the diaphragms were successively closed in steps
of 1\,mm. The data sets of Fig.~\ref{fig:2} (a,b,c) represent the transmission
probability for three different  values of the opening of the diaphragms
$w=5.8,\,8.8,\,$ and 15.8\,mm, respectively. Note the remarkable degree
of agreement between the measured and the calculated data without any
adjustable parameter. In Fig.~\ref{fig:2}a where $w/d \approx 0.37 $,
transport is suppressed and mediated only by resonance
scattering with narrow
Breit-Wigner shapes centered at the eigenenergies of the closed
billiard as indicated by the ticmarks. With increasing diaphragm
opening (Fig.~\ref{fig:2}b) transport acquires a significant
non-resonant contribution,
leading to the widening and the overlap of resonances. Finally,
for fully open leads (Fig.~\ref{fig:2}c), $w/d=1$ (or $s \approx 0$)
resonances appear as narrow window resonances in a non-resonant
continuum. The trajectory of the resonance parameter as a
function of $s$ can be both experimentally and theoretically
mapped out in considerable detail. Different types of resonances
can be identified by their characteristically different resonance
parameters. The evolution of the Fano parameter as a function of
$w/d$ (or $s$) for one resonance is highlighted in Fig.~\ref{fig:3}.
The transition from a narrow Breit-Wigner resonance via a
somewhat wider asymmetric Fano profile to a window resonance
is clearly observable. The good agreement with theory allows
to accurately determine the degree of decoherence present in
the experiment. As the Fano profile, in particular near its
minimum, is very sensitive to any non-interfering incoherent
background, we can determine  an upper bound for the damping
by comparison between experiment and theory to be
$\kappa \lesssim 10^{-4}$. As illustrated in Fig.~\ref{fig:3},
even a slightly larger value of $\kappa=10^{-3}$ 
would drastically deteriorate the agreement
between experiment and theory. In line with the value $\kappa=10^{-4}$,
we obtain an imaginary part of the complex Fano parameter for systems without
time-reversal symmetry \cite{brems} out of our fitting procedure as
$\rm{Im}\, q \lesssim 0.1$.
We note that by using superconducting cavities $\kappa$ could still 
be further reduced \cite{rich}, however with
little influence on the result, since we have already 
nearly reached the fully coherent
limit.\\Following the parametric evolution of a large number of resonances
we find a characteristic pattern of Fano resonance parameters
(Fig.~\ref{fig:4}). As example we show the evolution of the $q$-parameter
as a function of the opening $w/d$. Obviously, two distinct 
subsets of resonances appear in the
rectangular billiards: one set is characterized by a 
large and only weakly dependent asymmetry parameter $q$.
A second set of resonances features a strongly varying $q$ 
(on the log-scale!) from large values near the Breit-Wigner limit to values
close to $q \approx 0$ for wide opening, yielding a window
resonance. At the same time, the width $\Gamma$ first increases
with $w/d$ increasing from close to 0, then reaches a
local maximum and finally decreases slightly
when $w/d\rightarrow 1$ (not shown). A similar non-monotonic behavior
of $\Gamma$ was recently observed in a
single-electron transistor experiment \cite{goeres00}. Such
features can be understood in terms of avoided crossings
in the complex plane \cite{heiss91,burg95} between interacting
resonances. While the von Neumann-Wigner theorem \cite{neu29}
for bound states predicts avoided crossings between states of
the same symmetry and thus a non-monotonic variation of the
eigenenergy, interacting resonances can also display avoided
crossings on the imaginary axis \cite{heiss91,burg95}, i.e.~exchange
of the width of resonances and thus leading to a non-monotonic behavior
of one of the $\Gamma$ involved. The two resonance poles approach
each other in the complex energy plane and go through an avoided
crossing as a function of the coupling parameter $s$. As a result
of this ``resonance trapping'' effect,
the larger resonance gets even larger for increasing $s$ and will
form a background, on top of which the narrow resonance is situated 
\cite{ro}.\\
%This somewhat counterintuitive stabilization or narrowing 
%of the width of one resonance
%despite an increased opening, i.e.~in increased coupling to the
%environment, is sometimes referred to as ``resonance
%trapping'' \cite{ro,rott} and in the limit of $\Gamma \approx 0$
%as the ``formation of bound states in the continuum''. This 
%could possibly provide an alternative explanation for the results of
%Ref.~\onlinecite{goeres00}, where such a non-monotonic behavior was
%observed and has been previously explained in terms
%of increased impurity scattering \cite{brems}.
For the present system, the interacting resonances can be
completely characterized in terms of scattering wave functions
that can be unambiguously determined theoretically
(see inset Fig.~\ref{fig:4}).
Resonances that undergo a complete evolution from Breit-Wigner
resonances to a window resonance are all associated with the
{\it second} even excited state in the cavity, while resonances that
maintain their Breit-Wigner shape
are connected to transport through the transverse {\it ground state}
of the cavity. This mapping is controlled by the amplitude $p$
for transmission through the first transverse mode
[see Eqs.~(\ref{eqn:5},\ref{eqn:6})]. In the case that $p^2>1/2$
all resonances associated with the first mode are broader than
the resonances associated with the excited state and
vice versa for $p^2<1/2$. For geometric reasons
the scattering device studied here (Fig.~\ref{fig:1})
always favors transport through the first cavity mode and therefore
$p^2>1/2$. In this way we arrive at the remarkably simple result
that all resonances associated with a first mode feature 
a weakly varying $q$, while all
resonances associated with the second mode undergo the complete
evolution from the Breit-Wigner to the window limit. This one-to-one
mapping is supported by the data of Fig.~\ref{fig:2}, where only second-mode
resonances (indicated by the long ticmarks) ``survive'' the transition
of $w/d\rightarrow 1$, while all first-mode resonances (short ticmarks)
vanish in the background of the transmission spectrum.
The present observation has far-reaching implications for other systems.
By tracing the evolution of a given resonance as a function of a control
parameter the nature of the resonant channel can be uniquely determined.\\In 
summary, the rectangular microwave cavity attached to two
leads allows to study the interplay between resonant and
non-resonant transport in unprecedent detail. By controlled
change of the opening, tuning a Fano resonance from the Breit-Wigner
limit to the window resonance limit has become possible.
Fano resonances can be used to accurately
determine the degree of decoherence present in a scattering
device. Non-monotonic behavior of resonance parameters can
be related to avoided crossings between interacting resonances,
which can be unambiguously associated with different
resonant modes of the cavity. The latter feature is a
consequence of the separability of the wave function in the
closed cavity. Future investigations along these lines for non-separable
chaotic cavities promise new insights into the resonance dynamics
of open chaotic systems.

\begin{acknowledgements}
We thank D.-H.~Kim, C.~M\"uller, E.~Persson, I.~Rotter, C.~Stampfer,
and L.~Wirtz for helpful discussions. Support by the Austrian Science
Foundation (FWF-SFB016) is gratefully acknowledged.
\end{acknowledgements}

\begin{figure}[!tbh] 
\centering
\caption{(a) Schematic sketch of the rectangular cavity with leads
attached symmetrically on opposite sides. Exchangeable diaphragms
at the lead junctions allow to control the coupling between
the cavity and the leads. The open even transverse states are
indicated. (b) Photograph of the experimental setup.}
\label{fig:1}
\end{figure}

\begin{figure}[!tbh] 
\centering
\caption{Total transmission probability, 
$T^{\rm tot}(\sqrt{2\varepsilon}\,d/\pi,w/d)$, for 
transport through the rectangular cavity with three different 
openings of the diaphragms: 
(a) $w/d=37\%$, (b) $w/d=56\%$ and (c) $w/d=100\%$. 
For better comparison, the experimental and the calculated results are shown 
as mirror images. 
%For all values of $n$ shown above, two even transverse cavity modes 
%are excited whereas only one mode is open in the leads. 
The positions of all eigenstates
in the closed cavity are indicated by the gray ticmarks.
For all the calculated curves shown, a damping constant of $\kappa=10^{-4}$
was used.}
\label{fig:2}
\end{figure}

\begin{figure}[!tbh] 
\centering
\caption{Fano resonance near the 
second even excited transverse mode at $kd/\pi\approx 1.5095$. 
Experimental and theoretical result for four different cavity 
openings ($w/d$) are shown. Curves with equal $w/d$-ratio are displayed
in the same line style (solid, dashed, dotted, dash-dotted). For all 
calculated curves a damping factor 
$\kappa=10^{-4}$ was used, except for the additional gray dashed curve 
shown for which  $\kappa=10^{-3}$ and $w/d=0.68$.}
\label{fig:3}
\end{figure}

\begin{figure}[!tbh] 
\centering
\caption{Real part of the 
asymmetry parameter $|{\rm Re}(q)|$ as a function of the diaphragm 
opening $w/d$.
The shown data are obtained by fits to experimental data.
Solid circles ${\bullet}$ (open triangles $\triangle$) 
correspond to resonances originating
from the first (second) even cavity eigenstate. Typical wavefunctions 
$|\psi(x,y)|^2$ for these two classes of resonances are shown in the inset. 
The $\bullet$-resonances
always keep an $|{\rm Re}(q)|>10$, above
which the Fano resonances are very 
close to the Breit-Wigner lineshape [${\rm Re}(q)=\infty$]. 
The $\triangle$-resonances undergo a complete  
evolution from Breit-Wigner to window type as $w/d$ varies between 0 and 1.}
\label{fig:4}
\end{figure}

\end{document}